\documentclass[12pt]{article}

\def\fract_HP#1/#2{\leavevmode
  \kern.1em \raise .5ex \hbox{\the\scriptfont0 #1}%
  \kern-.1em $/$%
  \kern-.15em \lower .25ex \hbox{\the\scriptfont0 #2}%
}% 

\begin{document}
\title{\bf A vectorial form of the Schr\"odinger equation}

\author{Guy Barrand\thanks{barrand@lal.in2p3.fr}\\LAL, Univ Paris-Sud, IN2P3/CNRS, Orsay, France}
\date {June 2014}
\maketitle

\begin{abstract}
  We rewrite the time dependent Schr\"odinger equation by using
 only three dimensional vector algebra and by avoiding to introduce
 any complex numbers. We show that this equation leads to the same
 conclusions than the ``complex version'' concerning the hydrogen
 atom and the harmonic oscillator. We show also that this equation can be written as a Maxwell-Amp\`ere equation.
\end{abstract}



\section {The vectorial Schr\"odinger equation}
  Instead of writing the time dependent Schr\"odinger equation as
\begin{displaymath}
  i\hbar \frac{\partial \psi}{\partial t} = -\frac{\hbar^{2}}{2\mu}\Delta\psi+V\psi
\end{displaymath}
 we write it:
\begin{equation}\label{vec_sch}
  \hbar \frac{\partial }{\partial t} \{ \vec{\psi} \wedge \vec{\iota} \} =  -\frac{\hbar^{2}}{2\mu}\Delta\vec{\psi}+V\vec{\psi}
\end{equation}
with $\vec{\iota}(t,\vec{x})$ being a real unit three dimensional space direction
($\vec{\iota} \cdot \vec{\iota}=1$), $\vec{\psi}(t,\vec{x})$ being a three
dimensional real vector field and $\wedge$ being the vector wedge product. This equation does not contain complex numbers.

\section {The time independent vectorial Schr\"odinger equation}
 By taking a fix $\vec{\iota}$ we can recover the time independent Schr\"odinger equation with
 a $\vec{\psi}(t,\vec{x})$ real vector field of the form:
\begin{equation}\label{psi_1}
  \vec{\psi}(t,\vec{x}) = \exp(-\frac{Et}{\hbar} \vec{\iota} \cdot
  \vec{\omega})  \vec{\rho}(\vec{x})
\end{equation}
with $\vec{\rho}(\vec{x})$ being a real vector field orthogonal to $\vec{\iota}$
\begin{displaymath}
  \vec{\iota} \cdot \vec{\rho}(\vec{x}) = 0
\end{displaymath}
and $\vec{\omega}$ being the three matrices:
\begin{displaymath}
  (\omega_x)^j_k = \epsilon_{1jk} = \left(\begin{array}{ccc}0&0&0\\0&0&1\\0&-1&0\end{array}\right)
\end{displaymath}

\begin{displaymath}
  (\omega_y)^j_k = \epsilon_{2jk} = \left(\begin{array}{ccc}0&0&-1\\0&0&0\\1&0&0\end{array}\right)
\end{displaymath}

\begin{displaymath}
  (\omega_z)^j_k = \epsilon_{3jk} = \left(\begin{array}{ccc}0&1&0\\-1&0&0\\0&0&0\end{array}\right)
\end{displaymath}
with j,k,l = 1,2,3 and $\epsilon_{jkl}$ being the Levi-Civita
symbol. With all indices (\ref{psi_1}) looks:
\begin{displaymath}
   \psi^j = \left[\exp(-\frac{Et}{\hbar} \iota^l \omega^l)  \right]^j_k \rho^k
\end{displaymath}
\\
$\vec{\omega}$ is similar to $\vec{\sigma}$ formed
with the three Pauli matrices, and we have for example:
\begin{displaymath}
   ((\vec{v} \cdot \vec{\omega}) (\vec{w} \cdot \vec{\omega}))^j_k =
   w^jv^k - (\vec{v}\cdot\vec{w}) \delta^j_k 
\end{displaymath}
\begin{displaymath}
  (with \, \vec{\sigma} \, we \, have:\,
  ((\vec{v} \cdot \vec{\sigma}) (\vec{w} \cdot \vec{\sigma}))^j_k =
   i[(\vec{v} \wedge \vec{w}) \cdot \vec{\sigma}]^j_k +
   (\vec{v}\cdot\vec{w}) \delta^j_k \, )
\end{displaymath}
\\
As we can write:
\begin{displaymath}
  \frac{\partial}{\partial t} \{ \vec{\psi} \wedge \vec{\iota} \} = \frac{\partial \vec{\psi} }{\partial t}\wedge \vec{\iota} = (\vec{\iota}
  \cdot \vec{\omega}) \frac{\partial \vec{\psi} }{\partial t}
\end{displaymath}
\begin{displaymath}
  (with \, all \, indices \, it \, looks:
  \epsilon_{ljk} \frac{\partial \psi^j }{\partial t} \iota^k =
      \iota^k \epsilon_{klj} \frac{\partial \psi^j }{\partial t} =
      (\iota^k\omega^k)^l_j \frac{\partial \psi^j }{\partial t} \,)
\end{displaymath}
and:
\begin{displaymath}
  \frac{\partial \vec{\psi} }{\partial t} = -\frac{E}{\hbar}
  (\vec{\iota} \cdot \vec{\omega}) \vec{\psi}
\end{displaymath}
Since $\vec{\iota} \cdot \vec{\psi} = 0$ (because $\vec{\iota} \cdot \vec{\rho} =
0$) we have:
\begin{displaymath}
  \frac{\partial \vec{\psi} }{\partial t}\wedge \vec{\iota} = -\frac{E}{\hbar}
   ((\vec{\iota} \cdot \vec{\omega}) (\vec{\iota} \cdot \vec{\omega})) \vec{\psi}
  = -\frac{E}{\hbar} (\vec{\iota}(\vec{\iota}\cdot\vec{\psi})-\vec{\psi}) = \frac{E}{\hbar} \vec{\psi}
\end{displaymath}
  Then our vectorial equation (\ref{vec_sch}) reduces for this particular form (\ref{psi_1}) of
  $\vec{\psi}$ to a time independent Schr\"odinger equation over the vector
  field $\vec{\rho}(\vec{x})$:
\begin{displaymath}
  E \vec{\rho}(\vec{x}) = -\frac{\hbar^{2}}{2\mu}\Delta \vec{\rho}(\vec{x})+V \vec{\rho}(\vec{x})
\end{displaymath}

\subsection {Recovering the hydrogen atom}
 To recover the hydrogen atom, we seek a $\vec{\rho}(\vec{x})$ of the
 form:
\begin{displaymath}
  \vec{\rho}(r,\theta,\phi) = R(r,\theta)  \exp (m \phi \vec{\iota} \cdot
  \vec{\omega})  \vec{\kappa}
\end{displaymath}
 with $\vec{\kappa}$ being a fix unit vector orthogonal to
 $\vec{\iota}$ and $m$ being an integer number in
 order to recover the same $\vec{\rho}$ after a $2\pi$ rotation around $\vec{\iota}$ ($\vec{\rho}(r,\theta,\phi+2\pi) = \vec{\rho}(r,\theta,\phi)$). As we have:
\begin{displaymath}
  \frac{\partial^2 \vec{\rho} }{\partial^2 \phi} = -m^2\vec{\rho}
\end{displaymath}
  we can proceed as usual by using the Laplacian in spherical
  coordinates:
\begin{displaymath}
    \Delta \vec{\rho}= \frac{1}{r^2} \frac{\partial}{\partial r} ( r^2
    \frac{\partial \vec{\rho} }{\partial r})+
 \frac{1}{r^2\sin\theta} \frac{\partial}{\partial \theta} (
 \sin\theta \frac{\partial \vec{\rho} }{\partial \theta})+
 \frac{1}{r^2\sin^2\theta} \frac{\partial^2 \vec{\rho} }{\partial^2 \phi}
\end{displaymath}
 leading on $R(r,\theta)$ only to:
\begin{displaymath}
    \Delta R = \frac{1}{r^2} \frac{\partial}{\partial r} ( r^2 \frac{\partial R}{\partial r})+
 \frac{1}{r^2\sin\theta} \frac{\partial}{\partial \theta} (
 \sin\theta \frac{\partial R}{\partial \theta})-m^2\frac{R}{r^2\sin^2\theta}
\end{displaymath}
 and then to the traditional "$r,\theta$" Sch\"rodinger equation:
\begin{displaymath}
  -\frac{\hbar^2}{2\mu} \left[\frac{1}{r^2} \frac{\partial}{\partial r} ( r^2 \frac{\partial R}{\partial r})+
 \frac{1}{r^2\sin\theta} \frac{\partial}{\partial \theta} (
 \sin\theta \frac{\partial R}{\partial
   \theta})-m^2\frac{R}{r^2\sin^2\theta}\right]  +VR = ER
\end{displaymath}
  With a Coulombian potential:
\begin{displaymath}
   V = -\frac{e^2}{r}
\end{displaymath}
  we can proceed as usual; if E is quantised with:
\begin{displaymath}
   E_n = \frac{E_0}{n^2}
\end{displaymath}
 we have the solutions:
\begin{displaymath}
   R_{nlm}(r,\theta) = \rho_{nl}(r) Y_{lm}(\theta)
\end{displaymath}
\begin{displaymath}
   \vec{\rho}_{nlm}(r,\theta,\phi) = \rho_{nl}(r) Y_{lm}(\theta)  \exp (m \phi \vec{\iota} \cdot
  \vec{\omega})  \vec{\kappa}
\end{displaymath}
 and the real vector field solutions :
\begin{displaymath}
   \vec{\psi}_{nlm\vec{\iota}\vec{\kappa}}(t,r,\theta,\phi) = \rho_{nl}(r) Y_{lm}(\theta)  \exp(-\frac{E_nt}{\hbar} \vec{\iota} \cdot
  \vec{\omega})  \exp (m \phi \vec{\iota} \cdot
  \vec{\omega})  \vec{\kappa}
\end{displaymath}

\subsection {3D harmonic oscillator}
 In the same way, with the potential:
\begin{displaymath}
   V = \frac{1}{2} \mu \omega^2 r^2
\end{displaymath}
and the energy quantised as:
\begin{displaymath}
   E_{kl} = \hbar \omega (2k+l+\frac{3}{2})
\end{displaymath}
 we have the real vector field solutions :
\begin{displaymath}
   \vec{\psi}_{klm\vec{\iota}\vec{\kappa}}(t,r,\theta,\phi) = \rho_{kl}(r) Y_{lm}(\theta)  \exp(-\frac{E_{kl}t}{\hbar} \vec{\iota} \cdot
  \vec{\omega})  \exp (m \phi \vec{\iota} \cdot
  \vec{\omega})  \vec{\kappa}
\end{displaymath}

\section {Vectorial Schr\"odinger = Maxwell-Amp\`ere}
 If rewriting (\ref{vec_sch}) as:
\begin{equation}\label{vec_sch_xi}
  \partial_0 \{ \vec{\psi} \wedge \vec{\iota} \} = -\frac{\xi}{2} \Delta\vec{\psi} + {\cal V} \vec{\psi}
\end{equation}
with dimensions being $[\xi = \frac{\hbar}{\mu c}] = L$, $[{\cal V}] = 1/L$ and
$\partial_0 = \frac{\partial}{c \partial t} $. By using:
\begin{displaymath}
  \mathop{\rm rot} \mathop{\rm rot} \vec{\psi} = \mathop{\rm grad} \mathop{\rm div} \vec{\psi} - \Delta \vec{\psi}
\end{displaymath}
we can write:
\begin{equation}\label{sch_max}
  \mathop{\rm rot} \{ \frac{\xi}{2} \mathop{\rm rot} \vec{\psi} \} =
  \partial_0 \{ \vec{\psi} \wedge \vec{\iota} \} + \frac{\xi}{2} \mathop{\rm grad} \mathop{\rm div} \vec{\psi} - {\cal V} \vec{\psi}
\end{equation}
 By defining $\vec{H},\vec{D},\vec{j}$ as:
\begin{equation}\label{H_def}
  \vec{H} = \chi \frac{\xi}{2} \mathop{\rm rot} \vec{\psi}
\end{equation}
\begin{displaymath}
  \vec{D} = \chi \vec{\psi} \wedge \vec{\iota}
\end{displaymath}
\begin{displaymath}
  \frac{\vec{j}}{c} = \chi \{ \frac{\xi}{2} \mathop{\rm grad} \mathop{\rm div} \vec{\psi} - {\cal V} \vec{\psi} \}
\end{displaymath}
 ($\chi$ being a constant), we can write (\ref{sch_max}) as the Maxwell-Amp\`ere equation:
\begin{equation}\label{max_amp}
  \mathop{\rm rot} \vec{H} = \frac{\partial \vec{D} }{c\partial t} + \frac{\vec{j}}{c}
\end{equation}

\subsection {Other Maxwell equations}
 The form of (\ref{H_def}) strongly suggests to introduce $\vec{A},\vec{B}$ as:
\begin{displaymath}
  \vec{A} = \mu_0 \chi \frac{\xi}{2} \vec{\psi}
\end{displaymath}
\begin{equation}\label{B_H}
  \vec{B} = \mu_0 \vec{H} = \mathop{\rm rot} \vec{A}
\end{equation}
and then $U,\vec{E}$ as:
\begin{displaymath}
  \vec{E} = - \overrightarrow{\mathop{\rm grad}} U - \partial_0 \vec{A}
\end{displaymath}
 which, by construction, leads directly to the Maxwell-Faraday equation:
\begin{displaymath}
  \mathop{\rm rot} \vec{E} = -\frac{\partial \vec{B} }{c\partial t}
\end{displaymath}
and the Maxwell-Thomson equation:
\begin{displaymath}
  \mathop{\rm div} \vec{B} = 0
\end{displaymath}
 By defining $\rho$ as:
\begin{displaymath}
  \rho = \chi \mathop{\rm div} \{ \vec{\psi} \wedge \vec{\iota} \}
\end{displaymath}
 we have, by construction, the Maxwell-Gauss equation:
\begin{displaymath}
  \mathop{\rm div} \vec{D} = \rho
\end{displaymath}
 To be complete, we introduce the usual $\vec{M},\vec{P}$ vectors:
\begin{displaymath}
  \vec{H} = \frac{\vec{B}}{\mu_0}-\vec{M}
\end{displaymath}
\begin{displaymath}
  \vec{D} = \epsilon_0\vec{E}+\vec{P}
\end{displaymath}
We see that (\ref{B_H}) induces $\vec{M}=0$ and that $U,\vec{P}$ are related to
$\vec{\psi},\vec{\iota}$ with:
\begin{displaymath}
  -\epsilon_0 \overrightarrow{\mathop{\rm grad}} U + \vec{P} = \chi \{
  \vec{\psi} \wedge \vec{\iota} + \epsilon_0 \mu_0 \frac{\xi}{2} \partial_0
  \vec{\psi} \}
\end{displaymath}

\subsection {Dimensional analysis}
About dimensions, we have arranged to have:
\begin{displaymath}
  [\vec{D}] = [\vec{H}], [\vec{E}] = [\vec{B}] \Rightarrow [\epsilon_0 \mu_0] = 1
\end{displaymath}
By choosing the $\chi$ constant to be of dimension:
\begin{displaymath}
  [\chi] = \sqrt{energy [\epsilon_0]}
\end{displaymath}
and wanting to have the usual:
\begin{displaymath}
  [\vec{D}] = [\vec{H}] = \sqrt{ \frac{enery [\epsilon_0]}{volume} } \Leftrightarrow
  [\vec{E}] = [\vec{B}] = \sqrt{ \frac{enery}{volume  [\epsilon_0]} }
\end{displaymath}
then we get:
\begin{displaymath}
  [\vec{\psi}] = \frac{1}{\sqrt{volume}}
\end{displaymath}
which permit to stick to the fact that the integral:
\begin{displaymath}
  \int d^3x \vec{\psi} \cdot \vec{\psi}
\end{displaymath}
is without dimension.

\section {Currents}
In this paper we do not deal explicitly with the interpretation of $\vec{\psi}$ (and $\vec{\iota}$) but as
``charge densities'' and ``currents'' often help to shed lights on ideas we may
attach to mathematical symbols, we are going to consider this point
now.
\subsection{from (\ref{vec_sch_xi})}
 By taking the dot product of (\ref{vec_sch_xi}) with $\vec{\psi} \wedge \vec{\iota}$, we have:
\begin{equation}\label{vec_cont}
  \partial_0 \{ \left( \vec{\psi} \wedge \vec{\iota}
  \right) ^2\} + \xi \left( \vec{\psi} \wedge \vec{\iota} \right) \cdot \Delta
  \vec{\psi} = 0
\end{equation}
\begin{equation}\label{vec_cont_2}
  \Leftrightarrow \partial_0 \{ \vec{\psi} \cdot \vec{\psi} - \left(
      \vec{\psi} \cdot \vec{\iota} \right)^2
  \} + \xi \left( \vec{\psi} \wedge \vec{\iota} \right) \cdot \Delta
  \vec{\psi} = 0
\end{equation}
\textbf{With a constant $\vec{\iota}$}, by applying (\ref{vec_cont}) on a (still real) $\vec{\psi}$ of the form:
\begin{equation}\label{psi_2}
  \vec{\psi}(t,\vec{x}) = \exp \{ S(t,\vec{x}) \vec{\iota} \cdot
  \vec{\omega} \}  \vec{R}(t,\vec{x})
\end{equation}
some rather simple algebraic manipulations show that we have:
\begin{displaymath}
  \partial_0 \{ \vec{\psi} \cdot \vec{\psi} - \left( \vec{\psi} \cdot \vec{\iota} \right)^2\}+
  \xi \partial_k \{ \{ \vec{\psi} \cdot \vec{\psi} - \left( \vec{\psi}
    \cdot \vec{\iota} \right)^2 \} \partial_k S \} = -\xi \left(
    \vec{\psi} \wedge \vec{\iota} \right) \cdot \exp \{ S \vec{\iota} \cdot
  \vec{\omega} \}  \Delta \vec{R}
\end{displaymath}
which is a continuity equation that we rewrite:
\begin{equation}\label{psi_prob_cont}
  \partial_0 {\cal P} + \xi \vec{\partial} ( {\cal P} \vec{\partial} S ) = -\xi \left(
    \vec{\psi} \wedge \vec{\iota} \right) \cdot \exp \{ S \vec{\iota} \cdot
  \vec{\omega} \}  \Delta \vec{R}
\end{equation}
with $\cal P$ being then the positive density:
\begin{displaymath}
  {\cal P} = \vec{\psi} \cdot \vec{\psi} - \left( \vec{\psi} \cdot
    \vec{\iota} \right)^2 = \left( \vec{\psi} \wedge \vec{\iota}
  \right) ^2
\end{displaymath}
\\
  If imposing:
\begin{equation}\label{psi_3}
  \vec{R}(t,\vec{x}) = R(t,\vec{x}) \vec{\kappa}
\end{equation}
 with $\vec{\kappa}$ being a fix unit vector orthogonal to
 $\vec{\iota}$, then the right side term of (\ref{psi_prob_cont}) disappears and remains:
\begin{displaymath}
  \frac{\partial (R^2)}{c\partial t} + \xi \vec{\partial} ( R^2 \vec{\partial} S ) = 0
\end{displaymath}
which is the usual ``probability continuity'' equation found in
quantum mechanics text books. It is interesting to note that in the
case of the more general ``oriented'' $\vec{\psi}$ of the form
(\ref{psi_2}) the equation (\ref{psi_prob_cont}) does not reduce to
this last equation, but have anyway a positive density ${\cal P}$
which does not reduce to $\vec{\psi} \cdot \vec{\psi}$.

\subsection{from Maxwell}
By taking the divergence of (\ref{max_amp}), we have the conservation equation:
\begin{equation}\label{max_cont}
  \frac{\partial \rho}{\partial t} + \mathop{\rm div} \vec{j} = 0
\end{equation}
but with a $\rho$ being not compelled to be positive. Maxwell equations lead also to a
 continuity equation on a positive energy density:
\begin{equation}\label{max_ener}
  \partial_0 {\cal W} + \mathop{\rm div} \vec{\cal S} = -\vec{E} \cdot
  \{ \partial_0\vec{P} +\mathop{\rm rot} \vec{M} + \frac{\vec{j}}{c} \}
\end{equation}
with $\cal W$ and $\vec{\cal S}$ (Poynting vector) being:
\begin{displaymath}
  {\cal W} = \frac{1}{2} \{ \epsilon_0 \vec{E}^2 + \frac{\vec{B}^2}{\mu_0} \}
\end{displaymath}
\begin{displaymath}
  \vec{\cal S} = \frac{1}{\mu_0} \vec{E} \wedge \vec{B}
\end{displaymath}
Since we have:
\begin{displaymath}
  \frac{\vec{D}^2}{\epsilon_0} = \frac{\chi^2}{\epsilon_0} {\cal P} = \epsilon_0 \vec{E}^2 +...
\end{displaymath}
(\ref{psi_prob_cont}) is contained in (\ref{max_ener}) but does not
reduce to it. (If attempting to interpret things, we may say that the
quantum probability density ${\cal P}$ is contained, related to a Maxwellian energy!)

\section {$\vec{\iota}(t,\vec{x})$}
 The fact that we can write $\vec{\iota}(t,\vec{x})$, which makes no sense in case of using complex
 numbers (we can't do $i(t,\vec{x})$), introduces new modelling degrees
 of freedom which are not explored in this paper; but one idea could be
 that the $\vec{\iota}$ field be used as an ingredient for a
 ``collapse mechanism'' to explain why a system ``fixes'' to an
 eigen-vector-field $\vec{\psi_n}$ during a measurement or interaction. For the pleasure, we
 write again (\ref{vec_sch}) with all variables:
\begin{equation}\label{vec_sch_i_field}
  \hbar \frac{\partial}{\partial t} \{ \vec{\psi}(t,\vec{x}) \wedge
  \vec{\iota}(t,\vec{x}) \} =  -\frac{\hbar^{2}}{2\mu}\Delta\vec{\psi}(t,\vec{x})+V (t,\vec{x})\vec{\psi}(t,\vec{x})
\end{equation}

\section {Conclusions}
The equation (\ref{vec_sch}) without complex numbers is equivalent to
the Schr\"odinger one and it is remarquable that it leads to Maxwell
equations. Being able to extend ``$i$'' to a field $\vec{\iota}(t,\vec{x})$
leading to (\ref{vec_sch})  or (\ref{vec_sch_i_field}) opens modelling possibilities not explored in this paper.
The $i$ introduced by E.Schr\"odinger in 1925
in his equation had been the startup
in microphysics of an algebraic inflation that we find highly non
intuitive. (Would you claim to have a full intuitive understanding of all
symbols of a SUSY Lagrangian made with non commuting numbers? For our
point of view about quantum mechanics interpretations see \cite{ref-oo}). By avoiding
complex numbers in the grounding Schr\"odinger equation and using instead common vector geometry,
we hope to restore a more close relationship of intuition with
microphysics. But we have made only ``half the way'' since we do not
come yet with an interpretation of the $\vec{\psi}$ field which stay
unveiled. In particular we do not say if $\vec{\psi}$ is ontic or
epistemic, but three dimensional vector fields as $\vec{\psi}, \vec{\iota}$ would tend to an ontic
field interpretation.

\end{document}